\documentclass[a4paper,10pt]{article}
\usepackage{graphicx}
\usepackage{amsmath}
\usepackage[utf8]{inputenc}
\usepackage{hyperref}
\usepackage{indentfirst}
\usepackage{float}
\usepackage{subfig}
\begin{document}
 
\begin{center}
\mbox{}

{\large{\bfseries  {Collation of Feasible Solutions for Domain Based Problems: An Analysis of Sentiments Based on Codeathon Activity }}} 
\\[8pt]

\textbf{Rajeshwari K, Preetha S, Anitha C,  Lakshmi Shree K , Pronoy Roy }
\\[8pt]

*E-mail: {\tt rajeshwarik.ise@bmsce.ac.in, preetha.ise@bmsce.ac.in, anitha.bmsce@gmail.com, lakshmishreek@gmail.com, pronoyroy11@gmail.com }

\date{}
\end{center}
\vspace{0.2cm}

\begin{abstract}
    Codeathon activity is a practical approach for enduring the principles of Software Engineering and Object Oriented Modelling. Real world domain problem’s solution was accomplished through team work. Analysing the problem and designing a feasible solution through a one day activity was achieved through virtual connection. There are three different sections in a semester, 13 teams were framed and assigned one problem statement. Individual team were supposed to prototype a solution which was further used to build one feasible solution. The feedback from students showed different sentiments associated with day long activity. Vivid emotions and expressions of students were analysed. 
    \end{abstract}
    \textbf{}

\section{Introduction}
Teaching Learning Processes (TLP) is a well-planned methodology of achieving the learning objectives to meet the outcomes of a course. As part of TLP, codeathon was planned as an Alternate Assessment Tool (AAT). The activity was conducted virtually for the theoretical course Software Engineering and Object Oriented Modelling (SEO). Software Engineering course discusses the various approaches used in solving real world problems to build a software product or a service. Object Oriented Modeling emphasizes on the design and modeling of the solutions for software programmers. Codeathon activity was conducted by the department of ISE, B.M.S. College of Engineering (BMSCE), to analyse the sentiments of students in the age group of 20-22 years. The analysis was made on fifth semester students. BMSCE being an autonomous institute, gave faculties the liberty to structure the curriculum and framing new TLP for the courses. Students indulge in Activity-Based Learning more better rather than reading from prescribed textbooks. Codeathon was conducted to motivate students to explore their creativity and to understand the course practically.

In this practical approach, learning happens through activities and problem solving. Project-Based Learning is a technique where students learn through projects, acquire knowledge and develop skills by working together in a team. Faculties aimed at diverse teams of 3 or 4 students in a group, usually students prepare their teams and submit to course instructors. On a whole, the group consisted of a top performer, two average performers and a weak performer. Course teachers aimed at diverse groups with no friendships, where compatibility issues of knowing each other could be a concern. For a team, to know the strengths and weaknesses of the teammates would deliberately take more time than problem solving. Self-formed teams could drive the solutions or strategies quicker than a diverse team. Each team had to analyse and model the solutions for the given challenging real time problems. 

Codeathon activity was conducted to analyse and design a prototype for three domains \cite{narayan2021codeathon}. Students could further choose a sub-problem under the specific domain.
\\
The domains and the sub-problems were:
\begin {itemize}
\item
 Education Domain 
 \begin {itemize}
\item
 Conduction of Semester End Examinations (SEE) of Theory and practical sessions (On Campus examinations)
 \item
 Theory Classes and Lab experiments conduction. Classes to be held and how college should maintain norms with SOP of Karnataka Government during COVID-19 pandemic 
 \end{itemize}
 
\item
  Travel/Tourism Domain
  \begin {itemize}
\item
  How to enforce Government’s SOP during tourism and stay
 \item
   Travel (any mode of transportation) during COVID-19 pandemic 
 \end{itemize}
 
\item
   Agriculture Domain 
   \begin {itemize}
\item
   How technology can help farmers to bring Farm-to-Table \item
   Enabling Seed-to-Table revolution
   \end{itemize}
\end{itemize}

The teams had to understand, analyse the problem and work together to find unique design solutions.  For the evaluators, submissions were made in stages, to know the progress of each solution and design techniques. Each team had to submit manuscripts describing the design and solution models they create during each stage. Team members had to understand how codeathon activity works and how their concept can be transformed to prototype solutions rather than software development. Each team was also provided with three flags to take expert advice from the evaluators, which could compromise assessment marks. For the last two years, codeathon activity was conducted offline, due to pandemic COVID-19 it was conducted virtually. An analysis was made on the completion time and emotions of students to complete the task though distantly placed but, virtually connected. 

\section {Reference Section}
Dana et al. \cite{alzoubi2021teachactive} behavioural indicators of a student and instructor in a classroom can be focused on comparison of lower frequency of instructor talk with higher frequency of student conversing about a given assignment. Students are most likely to participate in groups for any classroom activity which is referred to as behavioural engagement. Rebecca \cite{achen2015evaluating} discusses engagement of students in classroom activities and analysing their viewpoint through a feedback survey. From the survey, the authors understood most students were benefited by the pedagogical approach as they get involved in the activities, rather than being a mute spectator. Marinko \cite{vskare2021impact} discusses the negative impact of the travel and tourism industry. Private and public sectors were hardly hit during the pandemic, causing an economic crisis across the globe. The tourism industry is also one of the sectors under stress which amplified financial cycles. The study  demonstrates suitable contingency plans to curb the effect of the pandemic, COVID-19. Shireen \cite{alshurideh2021factors} collected data samples of participants in the age group of 18-29 to know their usage of mobile as an examination tool. Mobile based services acted as a learning platform for connecting a teacher and a student virtually during the pandemic, COVID-19. The challenges faced by the student to attend virtual classes and online examinations were explored, due to connectivity issues and cloud services.

Venugopal K R \cite{mathapati2017sentiment} observed that sentiment analysis can be automated to know the sentiments of a user's comments. Natural language processing was used to know the opinions and to categorize the sentiments. For classification, extraction opinion words and target words were used to classify various dimensional data in heterogeneous domains. Kiet et al \cite{van2018uit} studied sentiment analysis on two major topics for classroom dynamics. The first task was to analyse the student feedback for a course teacher and the second task to analyse the feedback given by the course teacher to the students. Student given feedback was observed to be shorter for positive polarity and lengthier for the negative polarity. Most of the negative sentiments involved were appropriate for suggestions and reasoning for negativity.  Khin et al. \cite{aung2017sentiment} analysed the sentiments based on lexicon and machine learning techniques to classify the text. Table \ref {tab:my_label} includes additional key phrases added by the authors, which are not included in the Afinn Lexicon and Vader lexicon.
Daneena et al. \cite{dsouza2019sentimental} collected students’ feedback to analyse sentiments. The authors concluded that the Multinomial Naïve Bayes Classifier is more efficient than other methods like Random forests, NaïveBayes, Support Vector Machines. 
Eric et al. \cite{andersson2018methods} conducted a feedback to know the estimated hours of time spent by a student to complete any assignment, projects towards the course apart from regular class hours.  Also the authors collected the sentiments of the students to analyse the time spent in completing the extra workload.  The results showed neutral values as students had to compromise their vacation and to resume college activities. 
\begin{table}[ht]
    \centering
    \caption { Illustration of key phrases correlated with words \cite{aung2017sentiment}}
    \begin{tabular}{|l|l|}
    \hline
    \textbf{Type of Words} & \textbf {Key Phrases} \\
    \hline
   Negation & \textit {No, not, neither, nor, nothing, never, none} \\
    \hline
    Blind Negation & \textit{Need, needed, require, required} \\
    \hline
    Adjective, adverb, verb, noun words & \textit{Knowledgeable, interesting} \\
    \hline
    Intensifier word & \textit{Very, slightly, really}\\
    \hline
    \end{tabular}
    \label{tab:my_label}
\end{table}

\section{Codeathon Feedback Inferences}
Codeathon is a best-suited model to synthesize and evaluate students, an ideal platform to learn collaboratively and work to analyse the problem and showcase participants' skill sets. The activity’s goal was concerned with design and blueprints, rather than coding or implementation of the solution. This practical approach showcases the uniqueness of idealistic solutions and how students adhere to the timelines. The teamwork process emphasized on the inclusivity and diversity amongst the participants with varied proficiency and domain knowledge.

After the completion of the activity, students were asked to provide their feedback about the Codeathon event. A Google form was shared with some basic information to know their likes and emotions of the day-long activity.

The following were excerpts of the questionnaire:
\begin {itemize}
\item
Problem statement description
\begin {itemize}
\item
	Able to understand the problem statement
\item
	Informing the problem statement on the day of codeathon
\item
	Able to understand the problem, but not able to solve/design the solution
\end{itemize}
\item
Working in a team 
\begin {itemize}
\item
Would you perform better if you had a choice of selecting team members
\item
Was there any challenge in accepting the solutions discussed in the team
\end{itemize}
\item
Timelines and mode of working
\begin {itemize}
\item
Did you feel the work was tedious and hectic
\item
Did you feel the codeathon would have been more fun filled activity if conducted on campus
\item
Do you feel virtual interactions overloaded the activity, and had to rework on connectivity issues
\end{itemize}
\item
Evaluations and accomplishment of the solutions
\begin {itemize}
\item
	Did you feel to retain all your flags until last stage
\item
	Did you feel the inputs/suggestions given during the stage 1 evaluation helped you to focus/narrow down your theme
\item
Would you like to implement your solution, and give a product to the end users [to the college, or tourism department or to the farmer]
\end{itemize}
\end{itemize}

The pie charts depict the varied responses from students for the objective based questionnaire with interesting inferences. 
\begin {itemize}
\item
Students are more proactive when there is sprint mode for submissions 
\item
Irrespective of the team composition, students were more competent in problem solving
\item
Students preferred in-house codeathon rather than virtual meet. 
\item
Students wanted to defer the usage of flags till the end, and most of them restrained from using them as it would result in marks reduction.
\end{itemize}
\begin{figure}[H]
  \centering
  \includegraphics[width=0.6\textwidth]{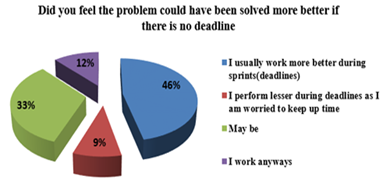}
\end{figure}
\begin{figure}[H]
  \centering
  \includegraphics[width=0.5\textwidth]{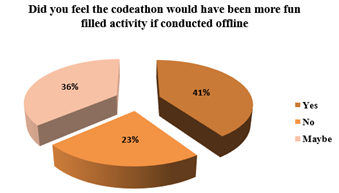}
\end{figure}
\begin{figure}[H]
  \centering
  \includegraphics[width=0.6\textwidth]{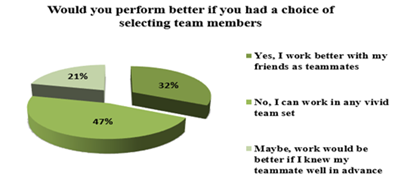}
\end{figure}
\begin{figure}[H]
  \centering
  \includegraphics[width=0.5\textwidth]{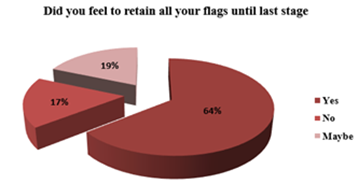}
\end{figure}

\section{Sentiment Analysis for the Codeathon Feedback}
The answers provided by participants of the survey to descriptive questions 
\begin {itemize}
\item
Was there any takeaway, will these activities similarly help you for other courses?
\item
Did this approach of practicality of the theory course improve your understanding of the concepts and inclination towards the same?
\end{itemize}
The feedback is pre-processed to remove punctuation's, special characters and other stop words. The Syuzhet package in R language is used to perform the sentiment analysis. It highlights various sentiments like trust, anger, anticipation, surprise, and depicts the measure of positivity and negativity associated with the sentiments of processed survey feedback data. The NRC Word-Emotion Lexicon comprises eight emotions; anger, fear, anticipation, trust, surprise, sadness, joy and disgust and two sentiments polarities; negative and positive. It contains a list of 14,182 words and their binary association scores with one of the eight emotions or one of the two polarities. In our study, the values obtained are scaled for the ease of comparison. To get the simple shape of the trajectory, Fourier Transformation (FT) and Discrete Cosine Transformation (DCT) are used. The feedback is categorized on the basis of chosen three domains Tourism/Travel, Agriculture/Farming and Education used in the codeathon and two important questions posed to the participants regarding the practical approach and prototyping were inquired

\section{ Results and Discussions}
\begin{figure}[H]
\centering
    \begin{minipage}[b]{0.45\linewidth}
        \includegraphics{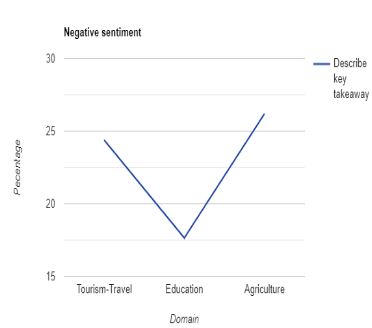}
        \caption{Negative sentiment variation for 'Describe key takeaway' across domains}
        \label{fig:1}
    \end{minipage}
    \quad
    \begin{minipage}[b]{0.45\linewidth}
        \includegraphics{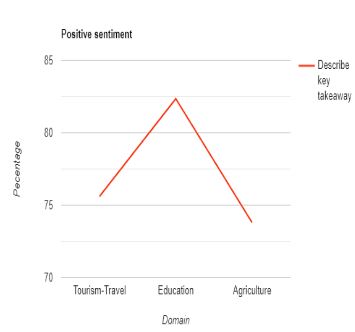}
        \caption{Positive sentiment variation for 'Describe key takeaway' across domains}
        \label{fig:2}
    \end{minipage}
\end{figure}

The variation of positive and negative sentiment for the answers provided by the participants to the question – ‘Describe your key takeaway from the activity’ has been shown in Figure \ref{fig:1} and Figure \ref{fig:2}. Out of the three domains, Education has the highest positive sentiment and Agriculture the lowest. The participants who were in the Agriculture domain displayed the highest negative sentiment which may be due to the lack of clear understanding of farm and agro-industry as the participants are from the urban setting. The highest positive sentiment for Education, Tourism-travel domain may be attributed to the clear understanding of students. Everyday problems and struggles to travel and attend classes in the university during the pandemic was the main concern.

\begin{figure}[H]
\centering
    \begin{minipage}[b]{0.45\linewidth}
        \includegraphics{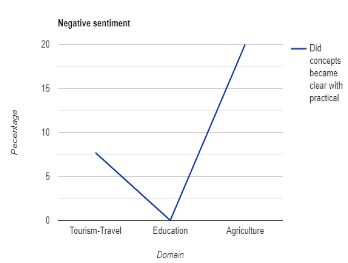}
        \caption{Negative sentiment variation for 'Did concepts become clear with practical}
        \label{fig:3}
    \end{minipage}
    \quad
    \begin{minipage}[b]{0.45\linewidth}
        \includegraphics{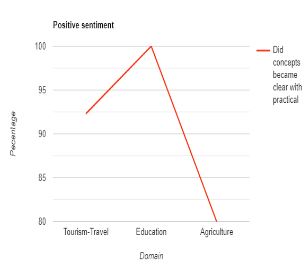}
        \caption{Positive sentiment variation for 'Did concepts become clear with practical’}
        \label{fig:4}
    \end{minipage}
\end{figure}

Further, Figure \ref{fig:3} and Figure \ref{fig:4} show the variation of polarity for answers to the question – ‘Did your practical knowledge improve’. Participants in the Education domain showed total positivity whereas; about 20\% of the participants in the Agriculture domain had negative sentiments. While some students felt the activity to be hectic within the given time constraints, some felt the need for better communication within team members to be able to perform better. Students felt the activity helped them to develop out-of-box-thinking, researching on a topic, helped focus on practical solutions to theoretical problems and improved team building skills.

\begin{figure}[H]
  \centering
  \includegraphics[width=0.4\textwidth]{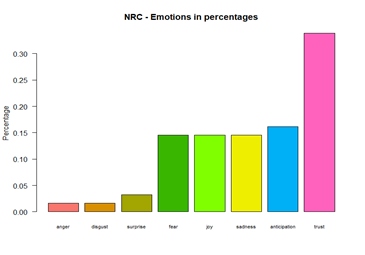}
  \renewcommand{\thefigure}{5}
  \caption{ Emotional variation for ‘Describe your key takeaway from the activity’ for Agriculture domain\label{fig:5}}
\end{figure}

\begin{figure}[H]
  \centering
  \includegraphics[width=0.4\textwidth]{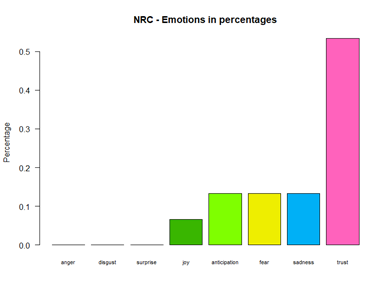}
  \renewcommand{\thefigure}{6}
  \caption{ Emotional variation for ‘Did your practical knowledge improve ?’ for Agriculture domain\label{fig:6}}
\end{figure}

\textbf{Emotional Analysis} \\
The variation of different emotions for the descriptive questions asked in the survey for the agriculture domain is shown in Figure \ref{fig:5} and Figure \ref{fig:6}. The answers provided by the participants in the Agriculture domain for both questions shows a high degree of trust. Further, it can be observed from Figure \ref{fig:5}, participants also expressed a high degree of anticipation from the evaluators for their problem domain. Students felt the need to resolve ambiguities, if any, towards seeking solutions. Very few participants also expressed anger in their key takeaway due to virtual meeting and connectivity issues.

\begin{figure}[H]
  \centering
  \includegraphics[width=0.4\textwidth]{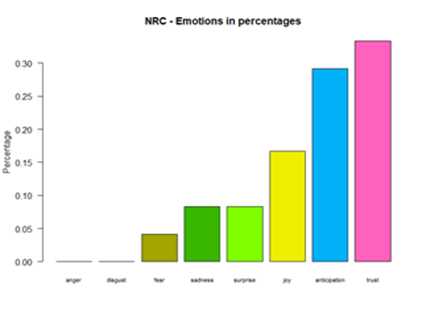}
  \renewcommand{\thefigure}{7}
  \caption{ Emotional variation for ‘Describe your key takeaway from the activity’ for Education domain\label{fig:7}}
\end{figure}

\begin{figure}[H]
  \centering
  \includegraphics[width=0.4\textwidth]{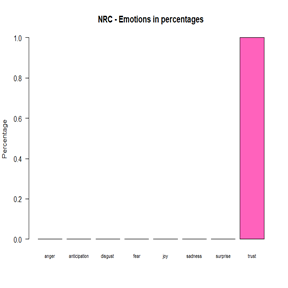}
  \renewcommand{\thefigure}{8}
  \caption{ Emotional variation for ‘Did your practical knowledge improve’ for Education domain\label{fig:8}}
\end{figure}

Trust was most expressed in feedback of participants of the Education domain as illustrated in Figure \ref{fig:7} and Figure \ref{fig:8}. Since participants were part of the system, they were able to connect to the problem statement better than the other two domains. Anticipation, joy and surprise were also the highest emotion after trust respectively in the feedback for key takeaway questions to the respondents. 

\begin{figure}[H]
  \centering
  \includegraphics[width=0.4\textwidth]{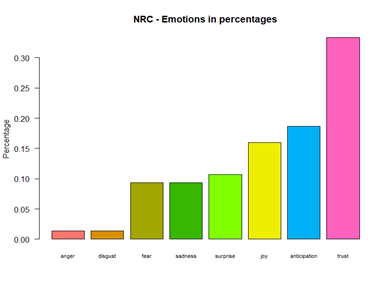}
  \renewcommand{\thefigure}{9}
  \caption{ Emotional variation for ‘Describe your key takeaway from the activity’ for Tourism-Travel domain\label{fig:9}}
\end{figure}

\begin{figure}[H]
  \centering
  \includegraphics[width=0.4\textwidth]{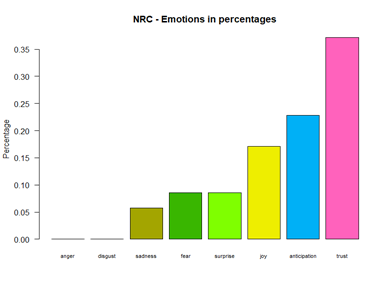}
  \renewcommand{\thefigure}{10}
  \caption{ Emotional variation for ‘Did your practical knowledge improve’ for Tourism-Travel domain\label{fig:10}}
\end{figure}

Similarly, participants in the tourism-travel domain also displayed trust the most, followed by anticipation and joy as shown in Figure \ref{fig:9} and Figure \ref{fig:10}. Anger and disgust were the least present emotion in both the answers for the domain. Surprise was present in more responses for ‘key takeaway’ questions as compared to ‘fear and sadness’. The emotional variation plots for all the three domains show that participants expressed trust and were joyous and surprised about the key takeaways from the event.

\begin{figure}[H]
 \centering
 \renewcommand{\thefigure}{11}
 \subfloat[Education]{%
      \includegraphics[width=0.25\textwidth]{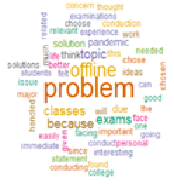}}
      \label{fig:11-a}
 \qquad
 \subfloat[Agriculture]{%
      \includegraphics[width=0.25\textwidth]{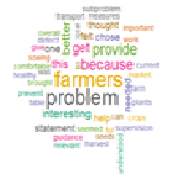}}
      \label{fig:11-b}
 \qquad
 \subfloat[Tourism-Travel]{%
      \includegraphics[width=0.25\textwidth]{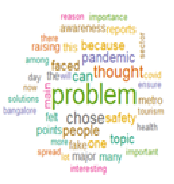}}
      \label{fig:11-c}
  \caption{ Word Cloud for 'why a sub-topic was selected' in a domain\label{fig:11}}
\end{figure}

The word clouds obtained from the responses provided to the question – “why a sub-topic was selected ?” by the participants in the survey is depicted in Figure \ref{fig:11}. The word ‘problem’ occurs with the highest frequency. This may be due to the reason that participants felt that problems relating to the domain due to pandemic needs to be addressed. For Agriculture, words like ‘farmers’, ‘help’, ‘transportation’, ‘market’ show that participants wanted to provide solutions for the problems faced by the farmers in transporting their produce to the markets. Further for the Education domain, words like ‘offline’, ‘exams’, ‘conducting’, ‘facing’ show that participants were keen to provide solutions to the problems faced by the students due to pandemic. Similarly, words like ‘safety’, ‘awareness’, ‘metro’, ‘COVID-19’ in the word cloud for Tourism-Travel show that the respondents of the survey were concerned about the safety and hygiene during travel.

\section{Conclusions}
The codeathon activity was conducted to converge multiple solutions of chosen three domains in Software Engineering and Object Oriented modeling, and how to collate feasible solutions obtained from the activity into one best solution for each problem. The emotional analysis highlights how difficult it is to obtain a best solution for a problem domain. Strangely, it was observed that the teams were not discussing the solutions amongst other teams of same domain unlike the in-house scenario. Virtual activity forced them to focus completely towards team connectivity and meeting the deadline and mere time to discuss with acquaintance. Evaluators picked the best solutions and converged them to a working prototype. Varied emotions of students were evaluated to understand their key takeaway from the one day activity. 

\bibliographystyle{unsrt}
\bibliography{mybib}

\begin{thebibliography}{10}

\bibitem{narayan2021codeathon}
Kausthub Narayan et~al.
\newblock Codeathon activity: A design prototype for real world problems.
\newblock {\em arXiv preprint arXiv:2107.10556}, 2021.

\bibitem{alzoubi2021teachactive}
Dana Alzoubi, Jameel Kelley, Evrim Baran, Stephen B.~Gilbert, Aliye
  Karabulut~Ilgu, and Shan Jiang.
\newblock Teachactive feedback dashboard: Using automated classroom analytics
  to visualize pedagogical strategies at a glance.
\newblock In {\em Extended Abstracts of the 2021 CHI Conference on Human
  Factors in Computing Systems}, pages 1--6, 2021.

\bibitem{achen2015evaluating}
Rebecca~M Achen and Angela Lumpkin.
\newblock Evaluating classroom time through systematic analysis and student
  feedback.
\newblock {\em International Journal for the Scholarship of Teaching and
  Learning}, 9(2):n2, 2015.

\bibitem{vskare2021impact}
Marinko {\v{S}}kare, Domingo~Riberio Soriano, and Ma{\l}gorzata
  Porada-Rocho{\'n}.
\newblock Impact of covid-19 on the travel and tourism industry.
\newblock {\em Technological Forecasting and Social Change}, 163:120469, 2021.

\bibitem{alshurideh2021factors}
Muhammad~Turki Alshurideh, Barween~Al Kurdi, Ahmad~Qasim AlHamad, Said~A
  Salloum, Shireen Alkurdi, Ahlam Dehghan, Mohammad Abuhashesh, Ra’ed
  Masa’deh, et~al.
\newblock Factors affecting the use of smart mobile examination platforms by
  universities’ postgraduate students during the covid 19 pandemic: an
  empirical study.
\newblock In {\em Informatics}, volume~8, page~32. Multidisciplinary Digital
  Publishing Institute, 2021.

\bibitem{mathapati2017sentiment}
Savitha Mathapati, SH~Manjula, et~al.
\newblock Sentiment analysis and opinion mining from social media: A review.
\newblock {\em Global Journal of Computer Science and Technology}, 2017.

\bibitem{van2018uit}
Kiet Van~Nguyen, Vu~Duc Nguyen, Phu~XV Nguyen, Tham~TH Truong, and Ngan
  Luu-Thuy Nguyen.
\newblock Uit-vsfc: Vietnamese students’ feedback corpus for sentiment
  analysis.
\newblock In {\em 2018 10th International Conference on Knowledge and Systems
  Engineering (KSE)}, pages 19--24. IEEE, 2018.

\bibitem{aung2017sentiment}
Khin~Zezawar Aung and Nyein~Nyein Myo.
\newblock Sentiment analysis of students' comment using lexicon based approach.
\newblock In {\em 2017 IEEE/ACIS 16th international conference on computer and
  information science (ICIS)}, pages 149--154. IEEE, 2017.

\bibitem{dsouza2019sentimental}
Daneena~Deeksha Dsouza, Divya P~Nayak Deepika, Elveera~Jenisha Machado, and
  ND~Adesh.
\newblock Sentimental analysis of student feedback using machine learning
  techniques.
\newblock {\em International Journal of Recent Technology and Engineering},
  8(14):986--991, 2019.

\bibitem{andersson2018methods}
Eric Andersson, Christopher Dryden, and Chirag Variawa.
\newblock Methods of applying machine learning to student feedback through
  clustering and sentiment analysis.
\newblock {\em Proceedings of the Canadian Engineering Education Association
  (CEEA)}, 2018.

\end{thebibliography}

\end{document}